\documentstyle[times]{plenum}
\input epsf

\begin{document}

\title{Colored Chaos}

\author{B. M\"uller \\ \bigskip
Department of Physics, Duke University,
Durham, NC 27708-0305}

\maketitle

\begin{abstract}

I review recent progress in our understanding of the basis of
statistical models for hadronic reactions and of the mechanisms
of thermalization in nonabelian gauge theories.

\end{abstract}

Almost to the date 30  years ago, Rolf Hagedorn proposed\cite{1}: that
multiparticle production and other phenomena of what today is called
``soft'' hadronic interactions could be explained on the basis of two
assumptions

\begin{enumerate}

\item The mass spectrum of hadronic states grows as $\rho(m) \sim
m^{-a}\;e^{bm}$.

\item The available states are statistically occupied during a (soft)
hadronic interaction.

\end{enumerate}

We now understand that the first assumption, an exponentially growing
mass spectrum, is a consequence of quark confinement, and the famous
Hagedorn temperature $T_H=1/b$ is related to the QCD string tension
and the temperature $T_c$ associated with the deconfining, chiral
symmetry restoring phase transition of QCD.

The second assumption has remained more mysterious.  Why is the
assumption of a random statistical distribution of final states
warranted in high energy reactions that last not much longer than 1
fm/c (or $3\times 10^{-24}$s), and why are the final states not
dominated by coherent quantum states or collective excitations of a
small subset of the available hadronic degrees of freedom?  Maybe it
is good to recall that similar questions posed themselves in the
context of Bohr's statistical model of compound nucleus reactions.  In
this case, the conceptual difficulties were eventually resolved by the
insight that the highly excited compound nucleus is a chaotic quantum
system \cite{2} exhibiting rapid exchange of energy between the accessible
degrees of freedom.\footnote{Experimental evidence for the chaotic
nature of the compound nucleus is mainly derived from the energy level
statistics of highly excited nuclei showing  the Wigner distribution
characteristic of chaotic quantum systems \cite{3}.} Here I want to show
that the same mechanism is responsible for the apparent thermalization in
high-energy hadron-hadron interactions:  nonabelian gauge theories are
strongly chaotic.

\section*{Chaos and Ergodicity}

\hspace*{\parindent}
A dynamical system exhibits {\em ergodic} behavior, if the time
average of an observable $A$ can be replaced by the phase space average
\begin{equation}
\lim_{T\to\infty} {1\over T} \int_0^T dt A(t) = \langle A\rangle
\equiv {1\over Z_E} \int d\Gamma_E A(\Gamma_E).
\end{equation}
$\langle A\rangle$ here denotes the {\em microcanonical} average,
$Z_E$ is the microcanonical partition function, and $d\Gamma_E$ is the
phase space measure at constant energy.  For systems with very many
degrees of freedom it is equivalent to take the {\em canonical} average
\begin{equation}
\langle A\rangle = {1\over Z(\beta)} \int d\Gamma A(\Gamma)
\exp[-\beta E(\Gamma)],
\end{equation}
where $Z(\beta) = \int d\Gamma \exp[-\beta E(\Gamma)]$ is the
canonical partition function and the inverse temperature $\beta=1/T$ is
determined by the condition $E = -\partial(\ln Z)/\partial\beta$.

For practical applications it is crucial to know the time scale on
which ergodicity is attained.  It can be shown that this time scale is
related to the rate $h$ of exponential divergence of neighboring
trajectories in phase space; this rate is called the (maximal)
Lyapunov exponent \cite{4}.  The complete spectrum of Lyapunov exponents is
defined as follows.  Consider a given trajectory in phase space,
$x_{\alpha}(t)$, where $\alpha = 1,\ldots\nu$ enumerates the degrees
of freedom.  $x_{\alpha}(t)$ is a solution of the classical equations
of motion of the system.  Now take a set of neighboring trajectories
(see Figure 1):
\begin{equation}
\tilde x_{\alpha}^{\prime(i)}(t) = x_{\alpha}(t) +
\delta x_{\alpha}^{(i)}(t),\quad \alpha = 1,\ldots,\nu.
\end{equation}

\begin{figure}

\def\epsfsize#1#2{0.8#1}
\centerline{\epsfbox{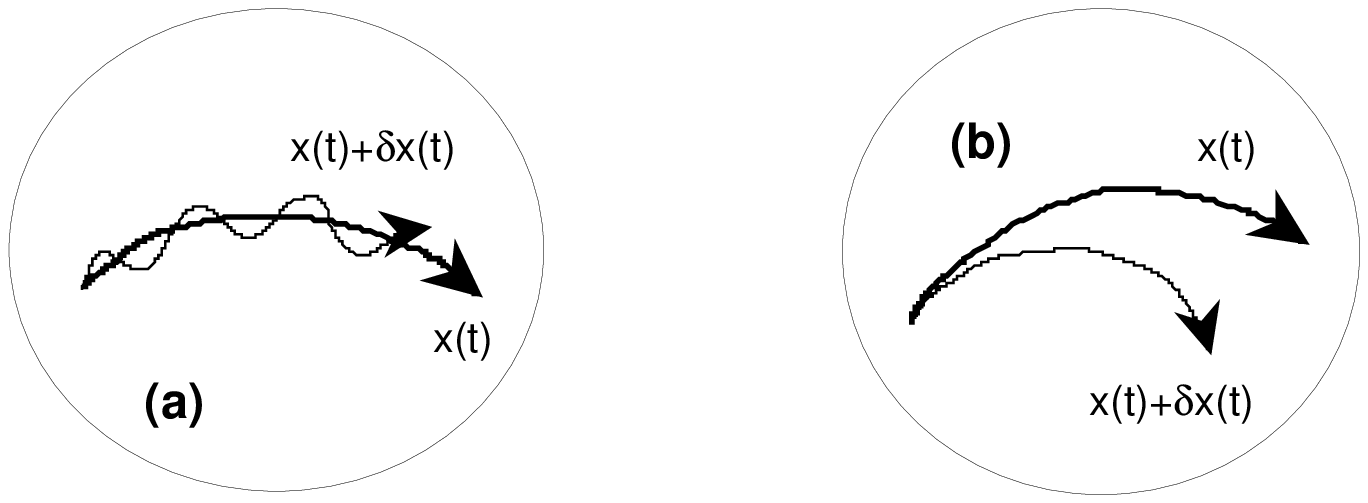}}

\caption{(a) Stable phase space trajectory;  (b) unstable trajectory.}

\end{figure}

For infinitesimal $\delta x_{\alpha}$ they are solutions of a
second-order linear differential equation of the form
\begin{equation}
D[x_{\alpha}(t)] \delta x_{\alpha}^{(i)}(t) = 0.
\end{equation}
One can then obtain a complete orthogonal set of solutions of this
equation; they define $\nu$ Lyapunov exponents $\lambda_i$ according to
\begin{equation}
\lambda_i = \lim_{t\to\infty} {1\over t} \ln {\Vert\delta
x_{\alpha}^{(i)}(t) \Vert \over \Vert\delta x_{\alpha}^{(i)} (0)\Vert},
\quad i=1,\ldots,\nu.
\end{equation}
In other words, for long times one has the norm of $\delta
x_{\alpha}^{(i)}$ growing (or shrinking) as $\Vert\delta
x_{\alpha}^{(i)}(t)\Vert \propto \exp(\lambda_it)$.  One usually
assumes the Lyapunov exponents to the ordered in size:
\begin{equation}
\lambda_1 \equiv h\ge \lambda_2 \ge \ldots \ge \lambda_{\nu}.
\end{equation}
For conservative (Hamiltonian) systems the Lyapunov exponents occur in
pairs of equal size, but opposite sign:  $\lambda_i =
-\lambda_{N+1-i}$.  This is in accordance with Liouville's theorem
which states that the volumes in phase space filled by an ensemble
remains unchanged with time, implying that there must be a direction
of contraction for every direction in which the phase space volume
expands.  For each conservation law there occur two vanishing Lyapunov
exponents; the conservation of energy always ensures the existence of
one such pair for a Hamiltonian system.  Since the extent of the
ensemble rapidly shrinks below any practially achievable resolution in
the exponentially contracting directions, the {\em observable} volume
in phase space grows as
\begin{equation}
\overline{\Gamma}(t) \propto \exp \left( t\sum_i
\lambda_i\theta(\lambda_i)\right) \equiv \exp (\dot S_{\rm KS}t)
\end{equation}
where the sum only includes the positive Lyapunov exponents.  The
exponential growth rate of the observable phase space volume implies
a {\em linear} rate of growth of the observable ``coarse-grained''
entropy associated with the ensemble.  This rate, $\dot S_{\rm KS} =
\sum_i\lambda_i\theta(\lambda_i)$, is called the Kolmogorov-Sinai
entropy, or short, KS-entropy.  Dynamical systems that have a positive
KS-entropy everywhere in phase space are called K-systems; they
exhibit all the properties required for a statistical description on
time scales that are long compared with the ratio between the
equilibrium entropy $S_{\rm eq}$ and the KS-entropy, i.e. for times
\begin{equation}
t \gg \tau_s = S_{\rm eq}/\dot S_{\rm KS}.
\end{equation}
An illustration of these properties is provided by the simple
dynamical system
\begin{equation}
H(x,y;p_xp_y) = \textstyle{{1\over 2}}\left( p_x^2+p_y^2+x^2y^2\right),
\end{equation}
which occurs as part of the extreme infrared limit of Yang-Mills
fields \cite{5}.  The system described by the Hamiltonian (8) has a positive
Lyapunov exponent $\lambda \approx 0.4$.  Almost all its trajectories are
unstable against small perturbations \cite{6} and the analogous quantum
system has been shown to exhibit a Wigner distribution of its level
spacings \cite{7}.  The remarkable ability of this system to randomize an
initially localized phase space distribution is shown in Figure 2.  After
a rather limited time the phase space distribution is indistinguishable
from a microcanonical ensemble.

\begin{figure}

\def\epsfsize#1#2{1.15#1}
\centerline{\epsfbox{72fig2.eps}}
\vskip-.5truein

\caption{Evolution of the phase space distribution for the model
Hamiltonian (9).  All points have energy $E=1$.  The apparent volume
of the phase space distribution grows rapidly, until it covers the
accessible phase space (contained within the hyperbolic boundaries)
homogeneously at time $t=32$.}

\end{figure}

\section*{Chaos in Nonabelian Gauge Theories}

\hspace*{\parindent}
If we want to apply these concepts to nonabelian gauge theories, we
must consider these as classical Hamiltonian systems with many degrees
of freedom, and we need a gauge invariant distance measure in the
space of field configurations.  The first part is easy; the
Hamiltonian formulation of lattice gauge theory by Kogut and Susskind
can form the basis for a study for the gauge group SU($n$) of nonabelian
gauge theories as dynamical systems.  The lattice Hamiltonian is
expressed as ($a$ denotes the lattice spacing)
\begin{equation}
H = {g^2\over 2a} \sum_{\ell} {\rm tr} (E_{\ell}E_{\ell}^{\dagger}) +
{2\over g^2a} \sum_p \left( n-{\rm Re}\; {\rm tr}\; U_p\right),
\end{equation}
where electric field strength $E_{\ell}$ and the link variables
$U_{\ell}$ are defined on the lattice links, and $U_p$ denotes the
ordered product of the $U_{\ell}$ around an elementary plaquette $p:
U_p=U_4^{\dagger}U_3^{\dagger}U_2U_1$.  The $U_{\ell}$ are elements of
the gauge group (SU(3) in the case of QCD) and the $E_{\ell}$ are
elements of the associated Lie algebra.  In the classical limit, the
link variables $U_{\ell}$ are functions of time, and the electric
field variables are given by
\begin{equation}
E_{\ell} = {a\over ig^2} \dot U_{\ell}U_{\ell}^{\dagger}.
\end{equation}
The Hamiltonian equations then provide a set of coupled equations for
the time evolution of $U_{\ell}(t)$ and $E_{\ell}(t)$, which can be
integrated numerically.  We have taken great care to ensure a numerically
exact solution.  The energy and Gauss' law remain conserved to better than
$10^{-6}$ over the whole course of the numerical integration.

An appropriate measure for the distance between two field
configurations is \cite{8}
\begin{equation}
D[U_{\ell},E_{\ell}; U'_{\ell},E'_{\ell}] = {1\over 2N_p} \left( a^2
\sum_{\ell}\left\vert {\rm tr} (E_{\ell}E_{\ell}^{\dagger}) - {\rm tr}
(E_{\ell}^{\prime}E_{\ell}^{\prime\dagger})\right\vert + \sum_p
\left\vert {\rm tr}\; U_p- {\rm tr}\; U'_p\right\vert\right).
\end{equation}
It is gauge invariant, gives a vanishing distance between gauge
equivalent field configurations, and goes over into
\begin{equation}
D[A,E; A',E'] \propto {1\over 2V} \int
d^3x \left( \vert E^2-E'^2\vert + \vert B^2-B'^2\vert\right)
\end{equation}
in the continuum limit, measuring the local differences in the
electric and magnetic field energy \footnote{If one only wants to
determine the largest Lyapunov exponent, it is sufficient to consider
either the electric or the magnetic contribution to $D[U,U']$.}.

If one starts from two randomly chosen neighboring field
configurations and integrates these in time, one finds that the
distance quickly grows exponentially, until it saturates due to the
compactness of the space of gauge fields.  The growth rate $h$ quickly
reaches a constant limit as function of the lattice size $N^3$, if the
energy density is kept fixed by choosing the same average energy per
plaquette $E_p$ in each case.  This is demonstrated in Figure 3 for
lattices of size $2^3$ up to $28^3$ and the gauge group SU(2).

\begin{figure}

\def\epsfsize#1#2{0.4#1}
\centerline{\epsfbox{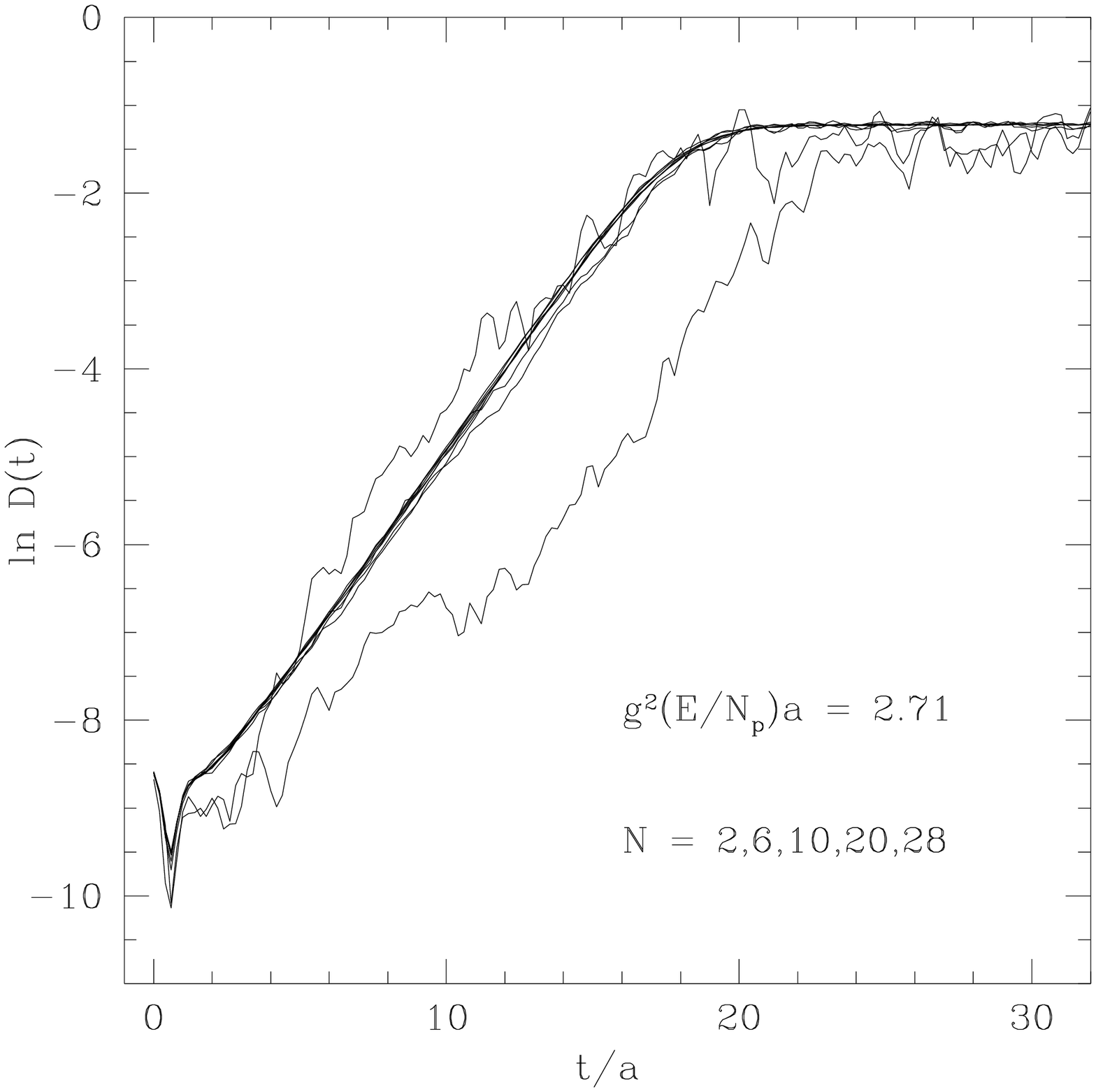}}

\caption{Growth of the logarithm of the distance between two randomly
chosen, initially neighboring, gauge field distributions.  The curves
are for $N^3$ lattices with $N=2,6,10,20,28$.  The fluctuating curves
are for $N=2$; rapid convergence occurs for larger $N$.  All curves
correspond to the same energy density.}

\end{figure}

It is easy to see\cite{9} that the Hamiltonian (9) exhibits a scaling
behavior such that the Lyapunov exponents, if they are universal
functions of the average energy density, as expressed by $E_p$, can only
depend on the dimensionless scaling variable $g^2E_pa$.  The nontrivial
surprise is that, as shown in Figs. 4a,b, the dependence is linear
\cite{8,10}
\begin{equation}
ha \equiv \lambda_1a = b_n g^2E_pa \quad \hbox{for SU($n$)},
\end{equation}
where $b_2\approx {1\over 6}$ and $b_3 \approx {1\over 10}$.  The
linear relationship means that the lattice spacing drops out, yielding
$h=b_ng^2E_p$ independent of $a$.  The maximal Lyapunov exponent hence
has a well-defined continuum limit.

\begin{figure}

\def\epsfsize#1#2{0.85#1}
\centerline{\epsfbox{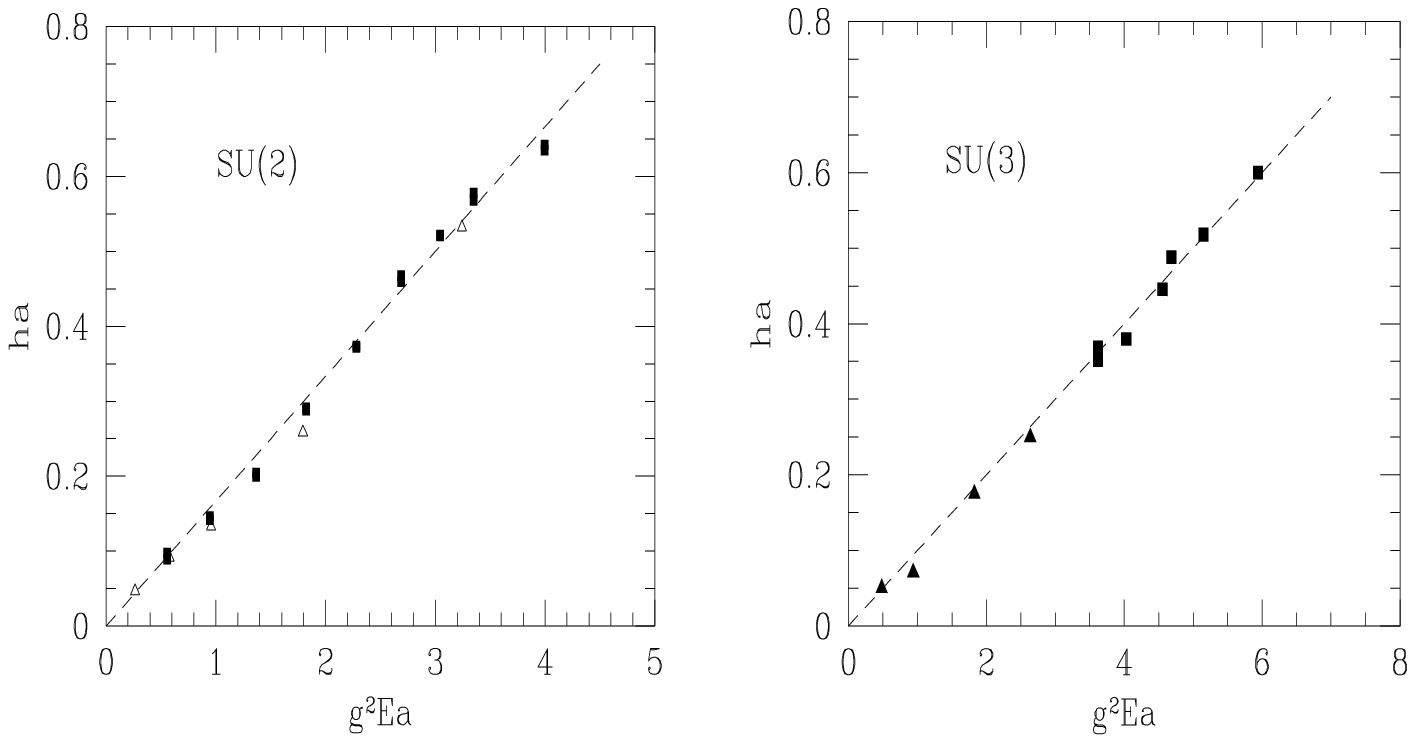}}

\caption{Maximal positive Lyapunov exponent as a function of the
scaling parameter $(g^2E_pa)$ for (a) SU(2), (b) SU(3) gauge theory.}

\end{figure}

What about the other Lyapunov exponents?  Their calculation for large
lattices is prohibitively expensive, as there are in total $6(n^2-1)N^3$
degrees of freedom for SU($n$) gauge theory on a $N^3$ lattice, but Gong
\cite{11} has evaluated the complete spectrum for SU(2) on lattices of size
$N=1,2$, and 3.  The result again is a surprise:  when the Lyapunov
exponents are scaled by the maximal one, and are plotted on the
interval $[0,1]$, the spectra for $N=2$ and $N=3$ are indistinguishable,
and there is only a small difference between $N=1$
and $N=2$ (see Figure 5).

\begin{figure}

\def\epsfsize#1#2{0.4#1}
\centerline{\epsfbox{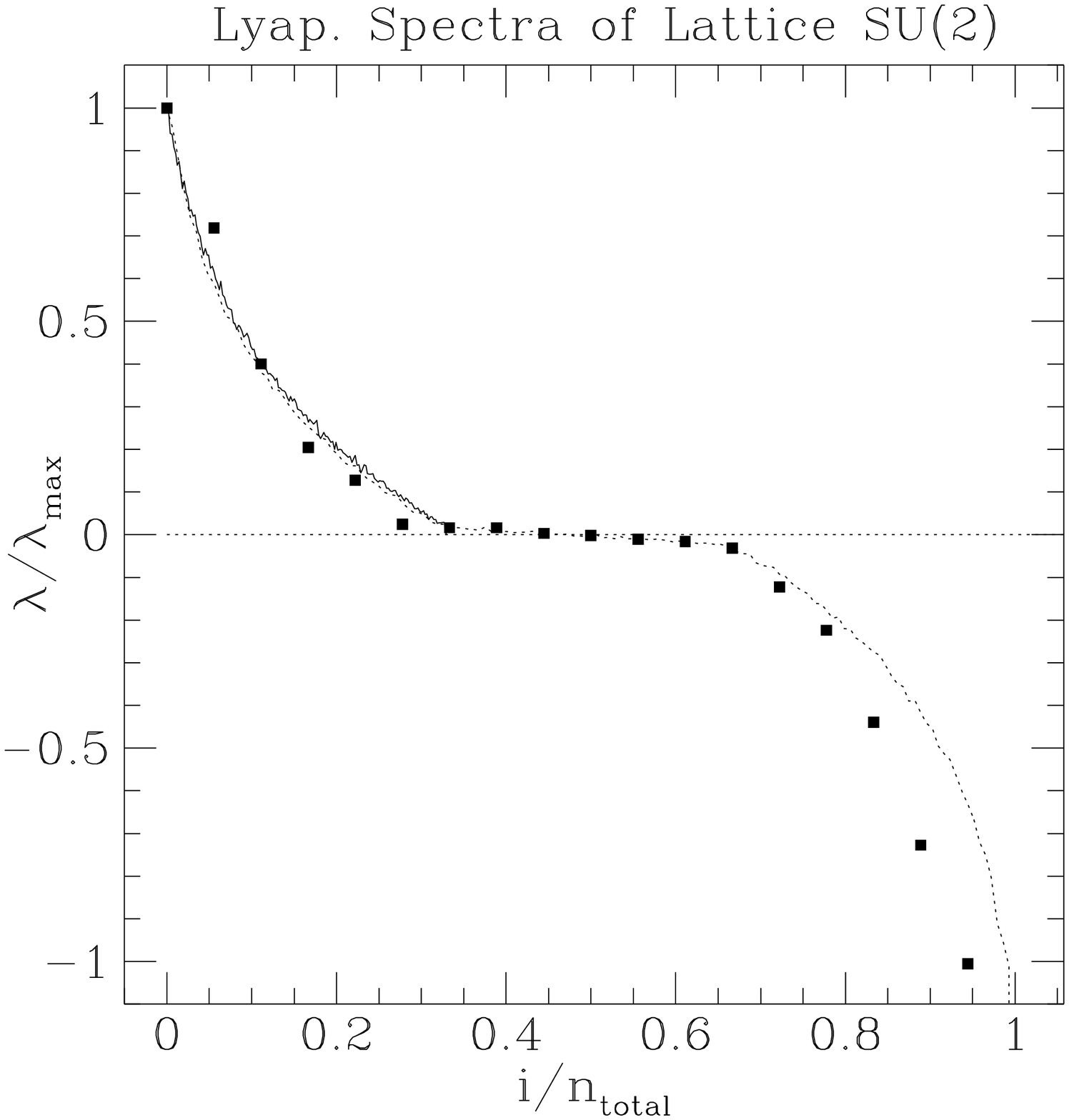}}

\caption{Spectrum of Lyapunov exponents for SU(2) lattice gauge
theory.  The black dots are for a $1^3$, the dotted line is for a
$2^3$, and the solid line is for a $3^3$ lattice.  The $18N^3$
exponents are plotted on the fixed interval [0,1] to exhibit the
scaling with $N$.}

\end{figure}

The Lyapunov spectrum for SU(2) shows three separate components:
there are $(6N^3-1)$ positive and negative exponents each, and there
are $(6N^3+2)$ exponents that converge to zero in the limit
$t\to\infty$.  Their vanishing reflects the existence of $(3N^3+1)$
conservation laws: Gauss' law at every lattice point and the overall
energy conservation.\footnote{The Lyapunov exponents associated with
Gauss' law obviously correspond to unphysical degrees of freedom and
only show up here because we did not fix the gauge explicitly in the
rescaling procedure used to determine the Lyapunov spectrum.  The fact
that they, indeed, vanish in the long-time limit provides support for
the numerical techniques employed in the calculation of the Lyapunov
spectrum.} Since the density of points on the line over the fixed
interval [0,1] grows as $N^3$, this implies that the sum over positive
Lyapunov exponents increases like the volume of the lattice, yielding
a constant {\em KS-entropy density} $\dot\sigma_{\rm KS}$ in the
thermodynamic limit.  Gong finds
\begin{equation}
\dot S_{\rm KS} = \sum_i \lambda_i\theta(\lambda_i) = c_2\lambda_1 N^3,
\end{equation}
which together with (14) and the lattice volume $V=(Na)^3$ yields
\begin{equation}
\dot\sigma_{\rm KS} = {1\over V} \dot S_{\rm KS} = {1\over 3}
b_2c_2g^2\varepsilon,
\end{equation}
where $\varepsilon=3E_p/a^3$ is the average energy density on the
lattice.  (Note that there are $3N^3$ plaquettes.)  No one has yet
calculated the complete Lyapunov spectrum for SU(3), but I expect a
similar relationship as (15) to hold in that case, too.  The
coefficient $c_2$ is not completely independent of the scaling
variable $g^2E_pa$, but has a value around 2.  We will return below to
the question how the physically relevant value of $g^2E_pa$ can be
chosen.

\section*{Physics Perspectives: Thermalization Time, Gluon Damping Rate}

\hspace*{\parindent}
The instability of all degrees of freedom of the nonabelian gauge
field (in the classical limit) leads to a very rapid
``thermalization'' of the energy density on the lattice.  This is
illustrated in Figure 6 showing the distribution of magnetic energy on
the lattice plaquettes \cite{12}.  The initial state was chosen according to a
random (not thermal) distribution of lattice link variables with
vanishing electric field everywhere.  Within two lattice units
($t/a=2$) the energy has been equilibrated between electric and
magnetic fields and, as the exponentially falling distribution shows,
has assumed the form of a Gibbs distribution.\footnote{I emphasize
that this ``thermalization'' is caused by the evolution of the gauge
field under its own Hamiltonian dynamics and not by some artifical
coupling to a heat bath as in the standard techniques applied in
Monte-Carlo simulations of lattice gauge theory.  There the
Monte-Carlo ``time steps'' have no physical meaning; here the time
step is physical.  The only approximation is that the lattice gauge
field is treated classically.} The time scale for this
``thermalization'' is in good agreement from the time scale estimated
from the inverse of the maximal Lyapunov exponent which is $h\approx
0.6$ in lattice units in SU(3) at this energy density.

\begin{figure}

\def\epsfsize#1#2{0.4#1}
\centerline{\epsfbox{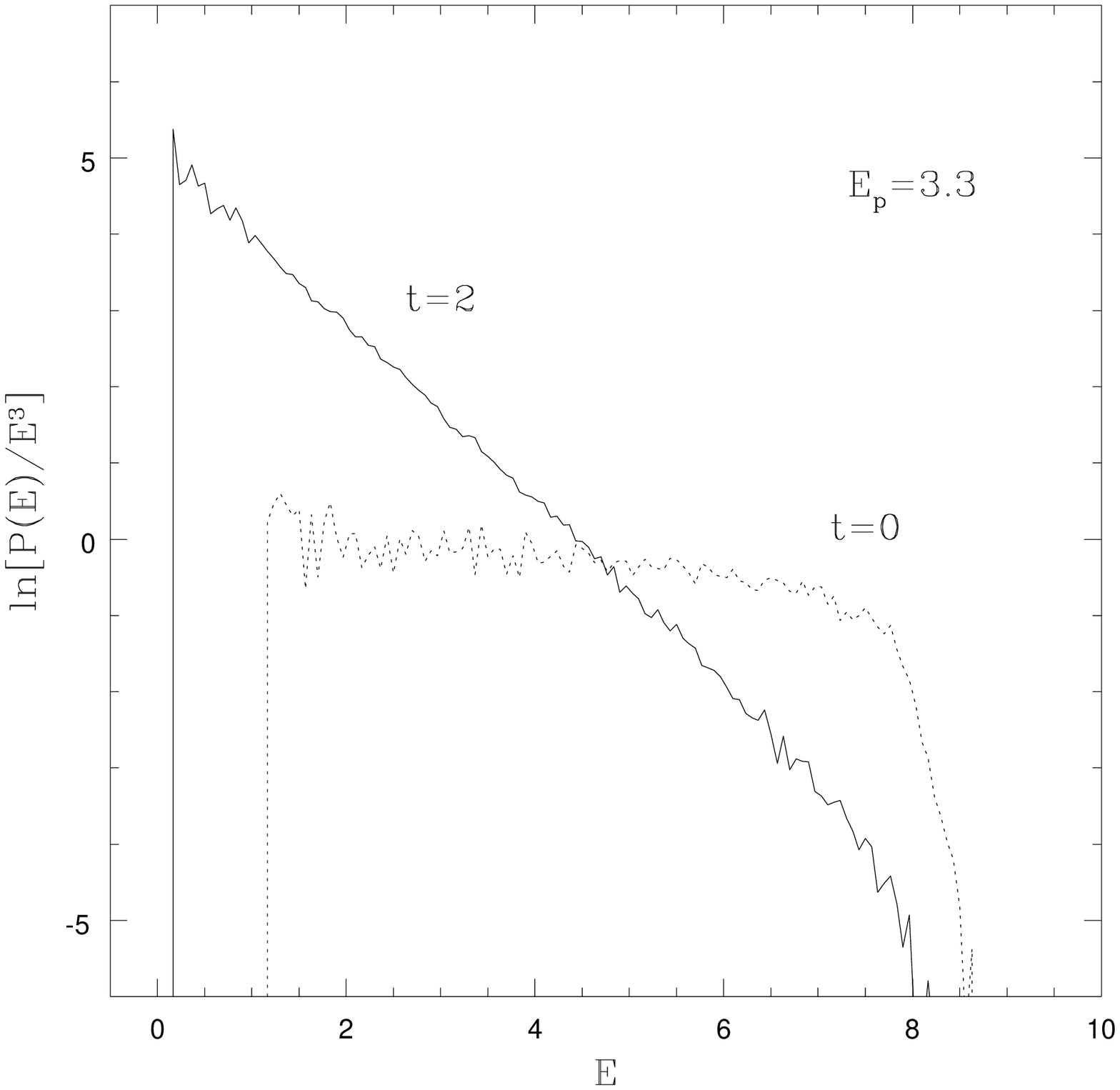}}

\caption{Logarithmic plot of the evolution of the magnetic energy density
distribution on the lattice for SU(3) gauge theory.  The distribution
appears ``thermalized'' (exponential) after a time $t\approx 2$.}

\end{figure}

The fact that the energy density thermalizes on a time scale much
shorter than that required for the numerical determination of the
Lyapunov exponents (typically $t/a= 1000$) allows us to relate the
Lyapunov exponents to quantities in the presence of a thermal
environment.  First, we can replace the average energy per plaquette
$E_p$ in (13) by the ``temperature'' $T$, because $E_p={2\over
3}(n^2-1)T$ for the classical, equilibrated SU($n$) gauge field.  This
implies that
\begin{equation}
h \approx \cases{{1\over 3} g^2T \approx 0.33 g^2T &for SU(2), \cr
\noalign{\medskip}
{8\over 15} g^2T \approx 0.53 g^2T &for SU(3). \cr}
\end{equation}
We can use this result to obtain a model independent, nonperturbative
estimate of the thermalization time solely due to gauge field dynamics
in QCD.  To compensate for the lack of ``running'' of the gauge
coupling constant in the context of our classical gauge field
calculation, we may use the one-loop result for $g(T)^2$ in (17) to
evaluate $\tau_h=h^{-1}$, as shown in Figure 7.  Clearly, this time is
much smaller than 0.5 fm/c for all relevant temperatures, indicating a
very rapid thermalization of the available energy.  One should note
that the Lyapunov exponents usually approach their asymptotic values
from {\it above}, i.e. the dynamical instabilities are actually
greater before the energy has been completely thermalized.  This
indicates that thermalization of field configurations far away from
equilibrium may proceed even more rapidly.

\begin{figure}

\def\epsfsize#1#2{0.55#1}
\centerline{\epsfbox{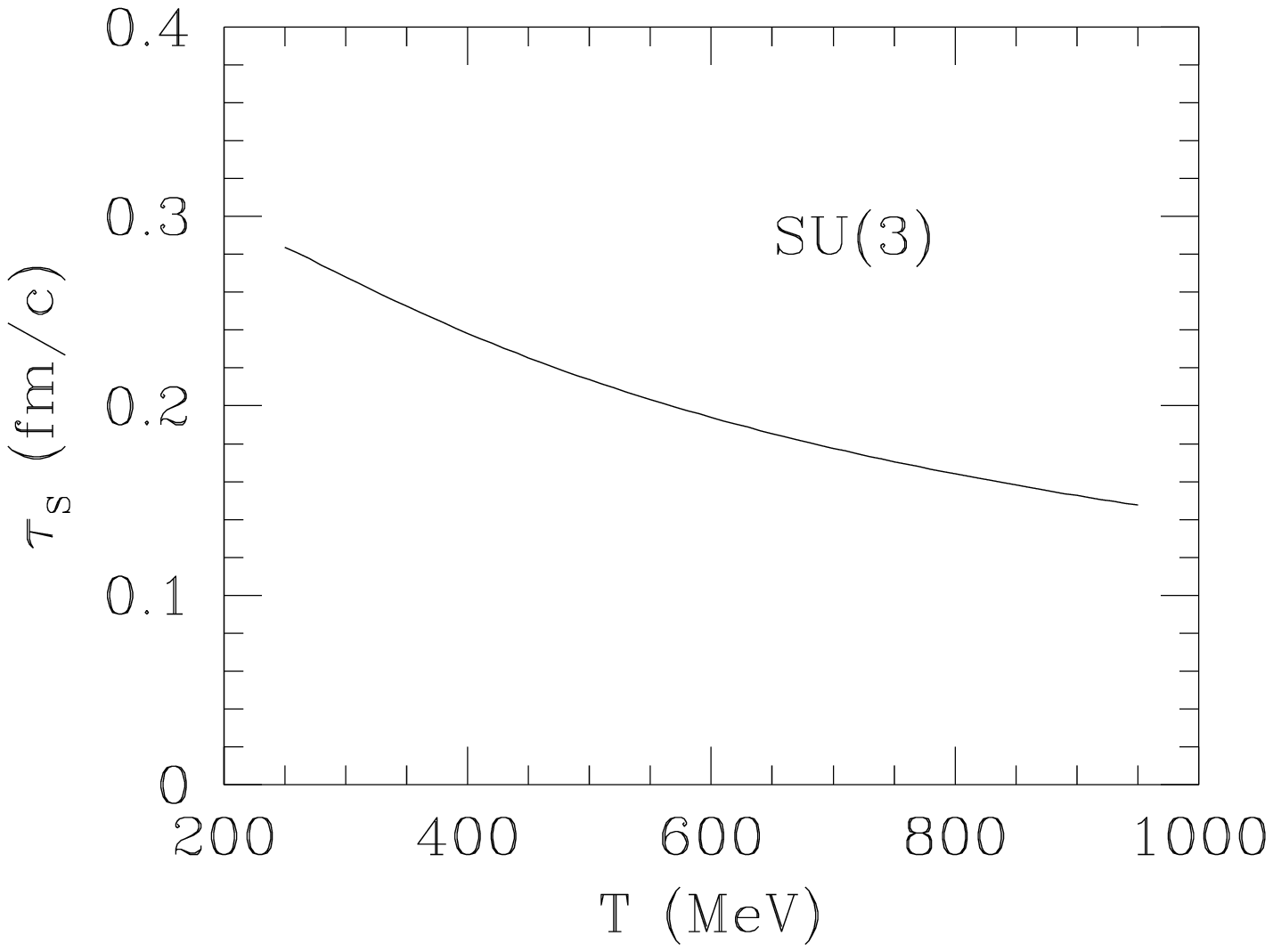}}

\caption{``Thermalization'' time scale in SU(3) gauge theory as
defined by the inverse of the maximal Lyapunov exponent (17),
using $g(T)^2 = 16\pi^2/[11\ln(\pi T/\Lambda)^2]$ with $\Lambda
= 200$ MeV.}

\end{figure}

It first appeared as a remarkable coincidence that the maximal
Lyapunov exponents for SU(2) and SU(3) agree within numerical errors
with the analytically calculated damping rate of a nonabelian plasmon
at rest \cite{13}:
\begin{equation}
\Gamma = 2\gamma (0) = 6.635 {n\over 24\pi} g^2T \approx \cases{ 0.35
g^2T &for SU(2), \cr 0.53 g^2T &for SU(3). \cr}
\end{equation}
[The reason for the factor 2 is that the plasmon pole is usally
parametrized as $\omega_p(k) = \omega^*(k)-i\gamma (k)$, so that the energy
density of the soft plasmon mode falls off as exp ($-2\gamma (0)t$).]
The observation that the Lyapunov exponent is numerically evaluated in
the vicinity of a ``thermalized'' field configuration over time scales
that are much longer than those of thermal fluctuations allows us to
establish a connection between these two quantities.  According to (4)
the Lyapunov exponents are determined from the long time behavior of
solutions of the linearized equation for the fluctuation $a_{\mu}(x,t)$
around an exact solution $A_{\mu}(x,t)$ of the Yang-Mills equations.
In the continuum limit this equation reads:
\begin{equation}
\left( g_{\mu\nu} D(A)^2 - D_{\mu}(A) D_{\nu}(A) \right) a^{\nu}(x,t)
-2i[F_{\mu\nu},a^{\nu}(x,t)] = 0
\end{equation}
where $D_{\mu}(A)$ denotes the gauge covariant derivative and
$[F_{\mu\nu},a^{\nu}]$ denotes the Lie algebra commutator.  Equation (19) is
the usual starting point for quantization in a background field, but
we will {\it not} impose here the background gauge constraint
$D_{\nu}a^{\nu}=0$.  The initial value problem for (19) can be solved
by means of the retarded propagator Schwinger function in the background
field.
\begin{equation}
a_{\mu}(x,t) = \int d^3x' \Delta_{\rm ret}^{\mu\nu} (x,t;x',0\vert A)
a_{\nu}(x',0).
\end{equation}
$\Delta_{\rm ret}^{\mu\nu}$ has the formal representation as difference
between the causal and the anticausal propagator:
\begin{equation}
\Delta_{\rm ret}^{\mu\nu}(A) = i\theta (t) \left[ D_{\rm F}^{\mu\nu}
(A) - D_{\rm F}^{\mu\nu^*}(A)\right]
\end{equation}
Now recall that the maximal Lyapunov exponent is obtained from the
long time average of the growth rate of $a^{\nu}(x,t)$.  Assuming
ergodicity, we may therefore replace the long-time average of the
propagators by the canonical, i.e. thermal, average:
\begin{equation}
\overline{\Delta}_{\rm ret}^{\mu\nu} = i\theta (t) \left[
D_{{\rm F},T}^{\mu\nu} - D_{{\rm F},T}^{\mu\nu^*}\right]
\end{equation}
where $D_{{\rm F},T}^{\mu\nu}$ denotes the {\it exact} finite
temperature Feynman propagator of the gauge field.  Note that the
causal Feynman propagator describes damped fluctuations, the
anti-causal propagator $D_{\rm F}^{\mu\nu}$ describes exponentially
growing perturbations.  The only remaining obstacle before establishing the
equivalence of $h$ and $2\gamma (0)$ is that the thermal average in
(22) should be calculated for a {\it classical} emsemble of gauge
fields, whereas the usual perturbative approach to $D_{{\rm F},T}^{\mu\nu}$
is based on the quantum mechanical ensemble.  However, this difference
turns out to be irrelevant for the plasmon damping rate $\gamma(0)$,
although it has a large effect on the effective plasmon mass
$\omega^*(0)$.  This is not entirely fortuitous, because the damping
rate is given by tree diagrams, such as Compton scattering and
bremsstrahlung, that have an exact low-energy classical limit
\cite{12,13}.  We can therefore identify the maximal Lyapunov exponent
with (twice) the damping rate of the most unstable mode in a thermal
nonabelianplasma, which turns out to be a plasmon at rest \cite{14}.

\section*{Conclusions and Outlook}

\hspace*{\parindent}
The short thermalization time scales of less than 1 fm/c found in our
studies of the time evolution of classical nonabelian gauge fields
show why Hagedorn was right thirty years ago with his assumption that
final states in ``soft'' strong interaction physics are populated
statistically.  The reason for the success of these classical studies
is that the dynamical instabilities in thermal gauge theories are of
order $g^2T$, which is a classical inverse length or time scale that
does not involve $\hbar$.  It would be interesting to see whether
other quantities of order $g^2T$, such as the thermal magnetic
screening mass on the ``spatial'' string tension, can also be
calculated in the framework of classical Yang-Mills theory.  This
immediately leads to the problem of deriving an effective
quasi-classical theory for thermal Yang-Mills theories at the length
scale $(g^2T)^{-1}$ that consistently incorporates quantum effects
from shorter distances in the form of transport coefficients.
Presumably such an effective theory will contain a gauge invariant
mass term of order $gT$ (as in the Taylor-Wong action) and a Langevin
noise term describing the fluctuations due to interactions with hard
thermal modes.

Another interesting problem concerns the application of real-time
evolution of gauge fields to processes far off equilibrium as they occur
in the earliest stage of hadron-hadron or nucleus-nucleus
interactions.  We have recently studied the instability of the
superposition of two counter-propagating plane waves, i.e. of a
standing abelian plane wave, in SU(2) Yang-Mills theory \cite{15}.
Here one finds that the Lyapunov exponent is proportional to the {\it
amplitude} of the wave, not to the energy as it is the case in random
fields.  Once the coherent wave is only slightly perturbed it decays
rapidly, exciting modes of all wavelengths, and quickly generates a
thermal energy spectrum (see Figure 8).  The evolution of more
realistic initial configurations, such as the interaction between
nonabelian wave packets, is presently under investigation.

\begin{figure}

\def\epsfsize#1#2{0.4#1}
\centerline{\epsfbox{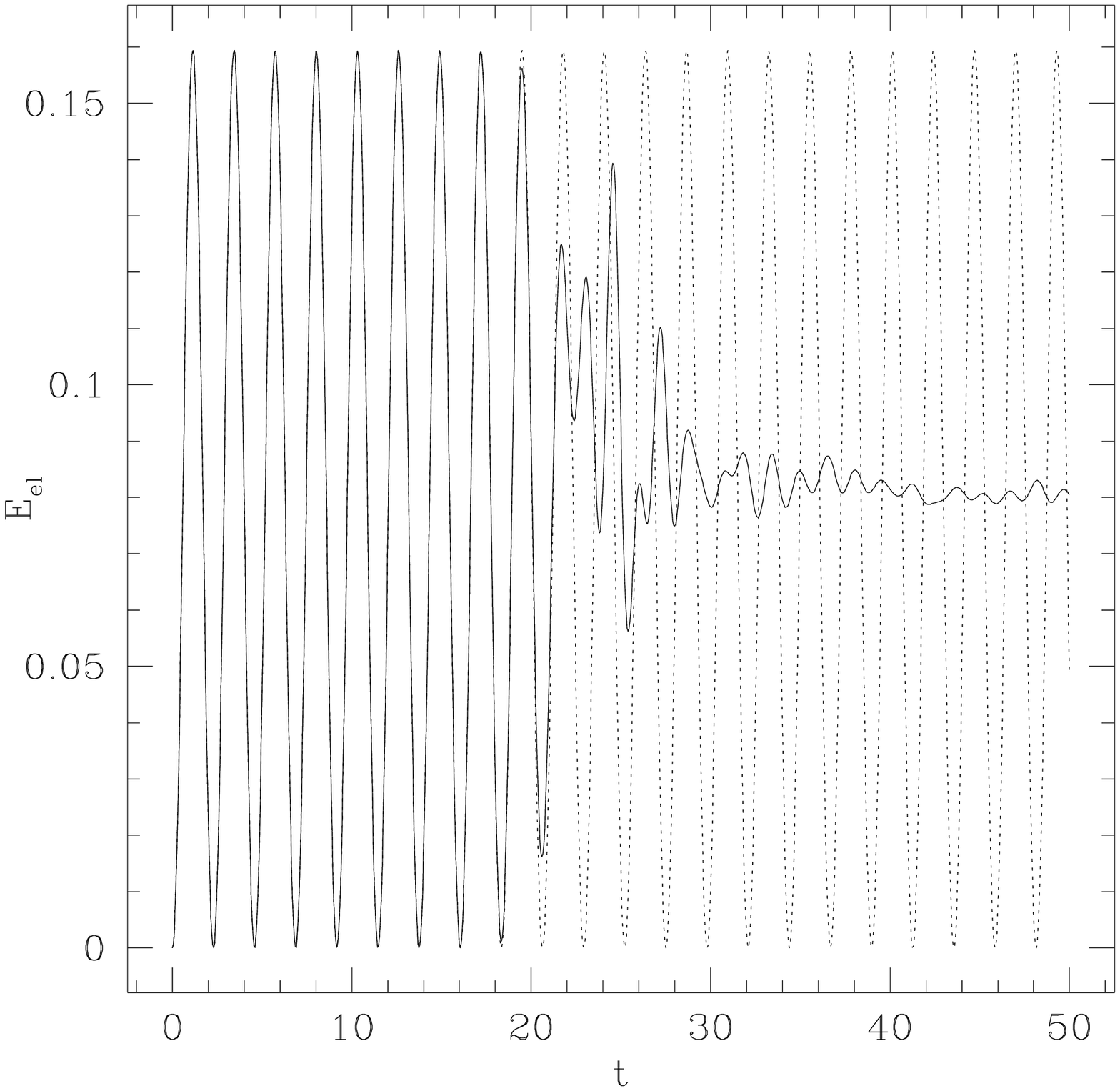}}

\caption{(a) Total electric energy for a SU(2) standing plane wave as
function of time.  The dotted curve shows the stability of an
initially abelian wave, the solid line shows the instability in the
presence of small nonabelian perturbations.  In that case the wave
eventually decays into a quasithermal frequency spectrum (b).}

\end{figure}

\section*{Acknowledgements}

\hspace*{\parindent}
I thank T. S. Bir\'o, C. Gong, S. G. Matinyan and A. Trayanov for
their enthusiastic help in unraveling the intricacies of chaotic
dynamics in gauge theories.  I would also like to thank D. Egolf, H.
B. Nielsen, S. E. Pugh, and G. K. Savvidy for illuminating discussions.  This
work was supported in part by the U. S. Department of Energy (grant
DE-FG05-90ER40592) and the North Carolina Supercomputing Program.

\vfill

\end{document}